%% file: main.tex
\documentclass[conference]{IEEEtran}
\usepackage{cite}
\usepackage{graphicx}
\usepackage{hyperref}
\usepackage[ruled,vlined]{algorithm2e}
\usepackage{qtree}
\usepackage{orcidlink}
\graphicspath{ {images/} }

\input{preamble}

\begin{document}

%Here goes the title

\title{Programming-By-Example by Programming-By-Example: Synthesis of Looping Programs}

%Authors List
\author{
    \IEEEauthorblockN{Shmuel Berman \orcidlink{0000-0002-6901-4852}}
    \IEEEauthorblockA{Columbia University\\
    New York\\
    s.berman@columbia.edu}
  \and
    \IEEEauthorblockN{Mark Santolucito \orcidlink{0000-0001-8646-4364}}
    \IEEEauthorblockA{Barnard College, Columbia University\\
    New York\\
    msantolu@barnard.edu}

}

\maketitle

%Main body starts

\begin{abstract}
\input{sections/0_abstract}

\end{abstract}

\begin {IEEEkeywords}
Program Synthesis, SyGuS
\end{IEEEkeywords}

\section{Introduction}
\label{intro}
\input{sections/1_introduction.tex}

\section{Motivating Example}
\label{section2}
\input{sections/2_motiv.tex}

\section{System Overview}
\label{section3}
\input{sections/overview.tex}

\section{Evaluation}
\label{section4}
\input{sections/eval.tex}

\section{Related Work}
\label{section5}
\input{sections/related.tex}

\section{Future Work}
\label{section6}
\input{sections/futurework.tex}

\section{Conclusion}
\label {conclusion}
\input{sections/conclusion.tex}

\bibliographystyle{IEEEtran}
\bibliography{main}

\end{document}

%% file: preamble.tex
\usepackage{xcolor}
\usepackage{xspace}
\usepackage{listings}

%% file: sections/0_abstract.tex
Program synthesis has seen many new applications in recent years, in large part thanks to the introduction of SyGuS. 
However, no existing SyGuS solvers have support for synthesizing recursive functions.
We introduce an multi-phase algorithm for the synthesis of recursive ``looplike'' programs in SyGuS for programming-by-example.
We solve constraints individually and treat them as ``unrolled`` examples of how a recursive program would behave, and solve for the generalized recursive solution. 
Our approach is modular and supports any SyGuS Solver.

%% file: sections/1_introduction.tex
%sygus is awesome
Syntax Guided Synthesis (SyGuS)~\cite{6679385} is a standardized format for program synthesis, which has made program synthesis widely accessible to end users and developers of synthesis-powered tools.
%however no lops
However, the programs that can currently be synthesized are limited to loop-free programs that simply process input data in a ``straight-line''. 
%loops are useful

Synthesizing looping and recursive programs is one of the major challenges to mainstream adoption of synthesis tools. Recursive programs are useful for modeling a variety of realistic synthesis scenarios and significant work has been done towards this goal.
%others have figured out loops
Madhusudan~\cite{madhusudan2011synthesizing} and Krogmeier Krogmeier et al.~\cite{krogmeier2020decidable} look at the synthesis of recursive programs by representing them as finite-length ASTs.
Humenberger and Kovacs~\cite{humenberger2021algebra} synthesized recurrence relations using polynomial invariants.
%but their solutions don't fit with sygus 
While these approaches have been successful, they do not fit the SyGuS synthesis model, meaning implementations cannot take advantage of the engineering advances of existing SyGuS solvers, such as CVC4~\cite{10.5555/2032305.2032319}.
%which is a problem, because sygus is so accessible, as we said before
%so we made something that does loops with sygus
While recursive programs are technically supported the by SyGuS standard, to the best of our knowledge, no major SyGuS solver supports synthesis of recursive program.

%how did we do it, from a high level theory perspective
To add support for synthesis of recursive functions in SyGuS, we propose an algorithm that breaks down SyGuS synthesis queries into simpler subproblems. 
In effect, we generate `examples' of how a recursive program should behave on differently sized inputs to synthesize that recursive program. 
Our insight is that while current SyGuS solvers cannot synthesize complete recursive/looping programs, solvers can find ``unrolled'' version of the program. 
We can then construct a new SyGuS query that combines the ``unrolled" subcomponents of the solution in order to synthesize a general looping program. 

%how did we implement it and evaluate it
We implement our modular procedure using the SyGuS solver CVC4~\cite{10.5555/2032305.2032319} as a black box. 
Our approach can be easily combined with any existing SyGuS solver. 
We provide a set of benchmarks that demonstrate our approach is able to solve SyGuS queries that require recursion.

In summary, our key contributions are:

\begin{enumerate}
    \item Provide examples of SyGuS queries (i.e. describing recursive functions) for which existing SyGuS solvers struggle to find concise solutions; 
    \item Propose an AST analysis technique for determining when a synthesized function is an ``unrolled`` recursive program;
    \item Propose a modular approach built on top of an existing SyGuS solver for synthesizing recursive programs;
    \item Implement our synthesis approach and provide an evaluation on a set of new SyGuS PBE benchmarks.
\end{enumerate}

%% file: sections/2_motiv.tex
Consider a user who wishes to synthesize a function, $f$, that repeatedly concatenates a string to itself according to the string's length. The  programming-by-example (PBE) constraints shown in Fig.~\ref{fig:exs} define such a function.

\begin{figure}[h!]
\centering

    {\renewcommand{\arraystretch}{1.3}
    {\small
    \begin{tabular}{|c c|} 
     \hline
     Input & Output  \\ [0.5ex] 
     \hline\hline
     synth & synthsynthsynthsynthsynth  \\ 
     \hline
     prog & progprogprogprog  \\
     \hline
     program & {\footnotesize programprogramprogramprogramprogramprogramprogram}  \\
     \hline
    \end{tabular}
    }
    }
    \caption{PBE Constraints}
    \label{fig:exs}
\end{figure}

With an adequate (non-recursive) grammar, a SyGuS solver may come up with a valid solution - however the solution will be large and incomprehensible (e.g. in this case, with AST size 79 when solved with CVC4).
However, this solution does not generalize to other inputs. 

To find the solution through our intuition, we would begin by finding a relationship between each input and its output; in this case, the first example is repeated five times, the second is repeated four, and the third is repeated seven times. 
An experienced programmer might then realized these are three ``unrolled'' recursive functions of repeated concatenation.

The programmer would then find a ``loop condition'' that uses features of the input to dictate how many times the string repeats.
In this example, the loop should run the same number of time as the string length.
To create a recursive solution, we create a function that concatenates a string \texttt{x} to itself \texttt{x.length} amount of times.

The code in Fig.~\ref{fig:code1} is the SyGuS solution generated by our tool. 
The AST of this solution is of size 18.

\begin{figure}[h!]
    \centering
\begin{verbatim}
(define-fun f 
  ((x String)) String 
  (g (x) (x) (- (str.len x) 1 )))

(define-fun g
  ((x String) (b String) (n Int)) String 
    (ite (<= n 0) 
      (b) (str.++ x (g (x) (b) (- n 1))))
\end{verbatim}
    
    \caption{Synthesized Functions}
    \label{fig:code1}
\end{figure}

%% file: sections/overview.tex
We divide our algorithm into five phases. We

\begin{enumerate}
    \item split the set of constraints into more easily solvable subsets. In our implementation, we place each query in its own subset individually. 
    \item pick a subset and run it through a SyGuS solver that can solve non-recursive queries (e.g. CVC4). We treat this solver as a black box.
    \item use a pattern detection module which detects ``loopable'' functions in the subset's solution. We then categorize the solutions based on their ``loopable'' function.
    \item iterate through each solution category and attempt to synthesize a loop condition.
    \item use a stitching module to compress the pattern(s) into a recursive solution and test them on every constraint. If any constraint fails, we return to Phase 2.
\end{enumerate}

%In this section, we first present a general pipeline for how we divide, compare, and unify the different constraints in parallel. We then offer an in-depth explanation for how we detect similarities in the solution structures, synthesize the loop condition based on those similarities. and then synthesize a generalized, loop-based solution.

%I have simplified the presentation of the SyGuS problem here. Having a clear definition and language of terms to use is important going forward.
In our algorithm we focus on programming-by-example (PBE).
In a PBE SyGuS query, we are given a grammar $G$ which defines the syntactic form of the solution, and a set of constraints $C$ in the form of input-output examples.
We write the set of input-output examples $C = \{ ex_1, ex_2, ... \}$, where each example $ex_i$ is a pair of values $(in_i, out_i)$.
The goal of synthesis is to find a program $f({x})$ such that $\forall ex_i \in C .\ f(in_i)=out_i$.

In Phase 1, we define a splitting module. 
Our goal in this step is to separate the constraints into the least amount of subsets possible, such that every constraint in a given subset requires an equal number of recursive calls. 
We do this because the more constraints a query has, the more specific the solution space is. 
Because we cannot (yet) predict how many recursive calls are required for each constraint, we place every constraint in a subset by itself, the subsets being $S_1={ex_1}, S_2={ex_2},.. S_n={ex_n}$. 

In Phase 2, we pick a constraint subset, $S_{i}$, and run it through a SyGuS synthesizer independently of all other constraint subsets using grammar $G$. 
It will generate a solution $f_i(x)$. 

In Phase 3, we start by finding out if we can express $f_i(x)$, or part of $f_i(x)$, as a repeated composition of a different function $g'(x)$. 

To find $g'(x)$, we first normalize the AST of solution $f_i(x)$. 
We then check each normalized AST for common contiguous subtrees. 
A subtree is common and contiguous if it has a child subtree that is identical to itself.

As a visual aid, we provide the following example AST, $AST_f$, representing the function $f'(x)$=$x+1+1+1$:

\begin{figure}[h!]
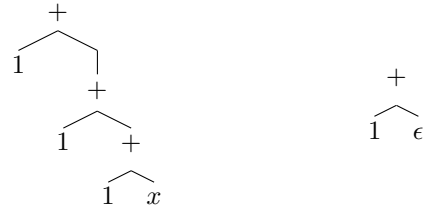

\begin{minipage}[.1\textheight]{.45\columnwidth}
 \Tree[.$+$ [.$1$
]
          [ [.$+$ [.$1$ ] 
                [.$+$ [.$1$ ]
                            [.$x$ 
                                 ]]]]] 
\label{fig:ast}
\end{minipage}
\begin{minipage}[.1\textheight]{.45\columnwidth}
 \Tree[.$+$ [.$1$
]
                                [.$\epsilon$ 
                                       ]] 
\end{minipage}
\caption{$AST_f$ and the detected common contiguous subtree $AST_g$}
\end{figure}

$AST_g$ is the repetitive subtree of $AST_f$, where $\epsilon$ is another copy of $AST_g$. 
If a common contiguous subtree is found in an AST, we count the number of repetitions and call this number $R_i$ (in Fig. 1, $R_i=2$). 

Once we have found $g'(x)$, we use it to categorize constraint subset $S_{i}$. Two constraint subsets belong to the same category $K_j$ if they are identical except in the number of times $g'(x)$ is composed with itself in the subset solution $f_i(x)$.

If we did not find a common contiguous subtree in $f_i(x)$, we do not place it in a category. If we did find one but it does not match with any existing category, we create a new category in their with constraint subset $S_{i}$ as its first element.

\iffalse
\begin{center}
 \begin{tabular}{||c c c||} 
 \hline
 Input & Output & $P(x)$  \\ [0.5ex] 
 \hline\hline
 $\alpha_1$ & $\beta_1$ & $T(S(S(S(Q(x)))))$  \\ 
 \hline
 $\alpha_2$ & $\beta_2$ & $T(S(S(S(S(S(Q(x)))))))$  \\
 \hline
 $\alpha_3$ & $\beta_3$ &  $T(S(S(S(S(Q(x))))))$\\
 \hline
\end{tabular}
\end{center}
\fi

We only enter Phase 4 if either a category has grown in size or a new category has been created in Phase 3.  Otherwise, we return to Phase 2. 

%If this condition is met,
In Phase 4, we take the PBE constraints in that category $K_j$ and create a set of new constraints out of their inputs and the corresponding $R_i$, the number of common contiguous sub-trees in solution AST $f_i(x)$. We use these constraints and grammar $G$ to synthesize a function whose output type is an integer. We run this query through the SyGuS solver to synthesize the ``loop'' condition $h(x)$. If this succeeds, we move to Phase 5.

In Phase 5, we create the recursive function and test it in on every constraint. We create a function $h'(x)$ from an arbitrary solution $f_j(x)$ in $K_j$, with all instances of the common contiguous subtree $AST_g$ replaced with a single placeholder node.
Our stitching module then transforms the function $g'(x)$, representing the common contiguous sub-tree $AST_g$, into a recursive helper function $g(x,in,n)$, where $n$ is the number of function repetitions and $in$ is the set of the original input variables.
%The variable $n$ represents how many times the function repeats and  $in$ is the set of the original input variables.
We then transform $h(x)$, our loop condition, and $h'(x)$, the non-recursive part of our solution, into $f(x)$. This function is identical to $h'(x)$ but calls $g(x,in,h(x))$ at the placeholder node.
As stated before, we then test this recursive solution on all constraints. If it succeeds, we return the solution. Otherwise, we return to Phase 2.

%% file: sections/eval.tex
As a preliminary evaluation, we created three string-manipulation PBE SyGuS benchmarks. We ran them using our tool and CVC4. We measure time to synthesize, and $|AST|$ size by counting nodes.

%\begin{figure}[h!]
%    \centering
\input{images/table}
%    \caption{Evalutation Table}
%    \label{fig:code}
%\end{figure}
These benchmarks were run on a machine with an Intel Core i7-8565U CPU @ 1.80GHz, 16GB 2400mHz DDR4 RAM running on WSL on Windows 10. We ran the benchmarks for up to 60 seconds.

Our modular algorithm is able to solve constraints that standalone CVC4 cannot. In cases when CVC4 can synthesize a solution, our algorithm is faster and produces a smaller, recursive solution.

Our implementation is available open-source at \url{https://github.com/shmublu/progsynth}.

%% file: images/table.tex
\begin{figure}[h!]
    \centering
\begin{center}
 \begin{tabular}{|c | c c | c c|} 
 \hline
  & \multicolumn{2}{c}{Our Tool} &  \multicolumn{2}{c|}{CVC4} \\
 Benchmark & Time & $|AST|$ &  Time  & $|AST|$ \\ [0.5ex] 
 \hline\hline
 1.sl~\cite{Berman1} & 2.432s & 20 & TO & - \\ 
 \hline
 2.sl~\cite{Berman2} & 2.467s & 24 & TO &  -  \\ 
 \hline
 3.sl~\cite{Berman3} & 2.455s &  37 & 6.387s & 214 \\ 
 \hline
\end{tabular}
\end{center}
    \caption{Benchmark Table}
    \label{fig:code}
\end{figure}

%% file: sections/related.tex
Our synthesis procedure enables the synthesis of recursive functions through a divide and conquer approach. 
Our approach can utilizes any existing SyGuS solver (in our implementation, CVC4) in a modular way.
Existing related work broadly falls into two categories; 1) work that also uses a divide and conquer approach to synthesis, and 2) work that has also tackled synthesis of recursive functions.
The existing divide-and-conquer approaches to synthesis have not been utilized for the synthesis of recursive functions, and the existing approaches to synthesizing recursive functions have not used SyGuS.

The main use of the divide-and-conquer approach in SyGuS have been utilized in helping SyGuS scale to larger synthesis problems.
Alur et al.~\cite{10.1007/978-3-662-54577-5_18} observed that SyGuS scales poorly as size of the minimal solution grows. 
To address this, Alur et al. use a divide-and-conquer approach whereby they find partial solutions that work for a subset constraints, and then synthesized predicates to stitch the partial solutions together.
Similar to our work, Alur et al. restrict themselves to a PBE specification model.
However their focus is on optimization rather than expanding the semantic search space of SyGuS to include recursive functions.

At the intersection of divide-and-conquer and recursive functions, Farzan et al.~\cite{farzan2021phased} proposes a tool, PARSYNTH, for synthesizing divide-and-conquer implementations of existing recursive algorithms.
PARSYNTH assumes that a recursive reference implementation already exists, whereas our goal is to synthesize that initial recursive implementation from a set of specifications.
Our work could be combined with PARSYNT as a post-processing step to have a divide-and-conquer algorithm that synthesizes divide-and-conquer programs.

In the space of synthesizing recursive programs, there has been a collection of works in this direction.
Generally these works have produced promising results, but are designed in a less modular way that our work, which builds on existing SyGuS solvers.
Generally, these works fall into three categories; taking inspiration from 1) reactive synthesis 2) type-driven synthesis and 3) SMT-solver driven synthesis.

Madhusudan~\cite{madhusudan2011synthesizing} works from a reactive synthesis perspective, synthesizing reactive recursive programs by representing them as finite-length ASTs. 
Their synthesis procedure uses reactive synthesis with specifications given in temporal logic, in contrast to our work which uses SyGuS with specifications given in a PBE style.
This work was later extended by Krogmeier et al.~\cite{krogmeier2020decidable} to give a decidable synthesis procedure for recursive programs (with some restrictions).
Again, although this line of work gives valuable insight to the problem of synthesis of recursive functions, it does not allow us to immediately leverage the prior successes of SyGuS solvers, like CVC4.

Feser et al.~\cite{10.1145/2737924.2737977} and Osera et al.~\cite{osera2015type} both focused on the synthesis of functional recursive programs using type-directed approaches.
Feser et al. synthesized functional programs using operations such as map and filter on data structures that are recursively expressed. Their procedure starts with general hypotheses and then synthesizes specific unknown programs, while we start from specific function behaviours and look for commonalities. Osera et al. also synthesized recursive functional programs by transforming the constraints into refinement trees. This enables them to prune the solution space by comparing constraints, which our approach in unable to do without voiding the possibility of a solution. Our algorithm can effectively only reduce the solution space if two constraints belong to the same category $K$, as we define in Sec.~\ref{section3}.

Thirdly, Kovacs and Humenberger.~\cite{humenberger2021algebra} synthesized loops and recursive functions, where their specification format was as a set of polynomial invariants.
In contrast, our work uses a PBE specification model and fits into the existing SyGuS synthesis model.

%% file: sections/futurework.tex
Currently, our algorithm supports basic applications of recursive or loop-like functions. 
Each loop through our synthesized function must perform the same set of operations, and as of yet we cannot vary the operation based on the loop iteration number. 
This stems from our common contiguous tree algorithm, where in order for two subtrees to be considered equal, both the structure and contents of the trees must be the same.
In order to support synthesis of programs where operations are parameterized by the loop iteration number, we would need to relax our definition of tree equality.
For two trees to potentially be captured by a loop operation, we only need the structure to be the same.
In some cases, the leaf nodes could be synthesized as ASTs as a function of the loop iteration number.

Additionally, due to our iterative approach of processing each constraint, the ordering of constraints can have an impact of running times. If a query is solvable by our tool, we can solve it regardless of the ordering, but the run time may change drastically. In an ideal ordering, we would sort constraints by the format of their individual solutions and then by descending order of how many times they recurse. We have not, as of yet, devised a method to do either of these tasks.

Our approach also requires the user to specify that they want to generate a recursive query ahead of time. Without this assumption, we would need to analyze the query and determine if it requires a recursive solution. Work by Morton et al.~\cite{morton2020grammar} used a neural network to detect which parts of the grammar were relevant in solving the query. A similar approach could be utilized in detecting if a query requires a recursive solution.

%% file: sections/conclusion.tex
Our approach's potency is derived from the power of the underlying SyGuS solver. If the SyGuS solver can synthesize the loop condition $g'(x)$ and individual solutions $f_i(x)$, then we can extend its capabilities to synthesize fully recursive functions.

We present a modular tool that implements this approach while adding negligible time to the existing underlying SyGuS solver. Our approach is scalable to support many more categories of recursive functions, and is useful in solving realistic problems.